# Reconstruction of cosmological density and velocity fields in the Lagrangian Zel'dovich Approximation

Rupert A.C. Croft[1] and Enrique Gaztañaga[2,3]
[1] *Department of Astronomy, The Ohio State University, Columbus, Ohio 43210, USA*
[2] *Research Unit (CSIC), Institut d'Estudis Espacials de Catalunyaq (IEEC), Edf. Nexus-104 - c/ Gran Capitan 2-4, 08034 Barcelona*
[3] *Oxford University, Astrophysics, Keble Road, Oxford OX1 3RH, UK*

**ABSTRACT**

We present a method for reconstructing cosmological density and velocity fields using the Lagrangian Zel'dovich formalism. The method involves finding the least action solution for straight line particle paths in an evolving density field. Our starting point is the final, evolved density, so that we are in effect carrying out the standard Zel'dovich Approximation based process in reverse. Using a simple numerical algorithm we are able to minimise the action for the trajectories of several million particles. We apply our method to the evolved density taken from N-body simulations of different cold dark matter dominated universes, testing both the prediction for the present day velocity field and for the initial density field. The method is easy to apply, reproduces the accuracy of the forward Zel'dovich Approximation, and also works directly in redshift space with minimal modification.

## 1 INTRODUCTION

Structure in the present day Universe is believed to have been formed as a result of the growth of small amplitude energy-density perturbations in the early Universe via the mechanism of gravitational instability. On scales significantly larger than individual galaxies we expect gravity to be the dominant effect so that we can use gravitational dynamics to relate theoretical predictions for the initial fluctuations with the present day observed Universe. These observations are mainly of the spatial distribution of galaxies, from redshift surveys (e.g. Davis & Peebles 1983, Fisher et al. 1994).

With the above assumption, we can in principle use our knowledge of gravity to recover the initial conditions and compare with theories for the generation of initial fluctuations, such as inflation which predicts a Gaussian distribution of Fourier amplitudes. In this paper, we present a simple method for doing this, which uses as its starting point the present day positions of galaxies in redshift space.

Prior to a redshift of $z \sim 10^3$, the nature of dark matter will have been important in the evolution of fluctuations and this will also be reflected in the initial conditions Direct measurements of galaxy peculiar velocities (Dressler et al. 1987, Burstein 1990, Willick et al. 1995) and comparisons with predictions of gravitational instability can provide an important consistency check on the above procedure.

Methods for determining the gravitational evolution of cosmological density perturbations can be split into two categories - the Eulerian and Lagrangian approaches. In the Eulerian approach (see e.g. Lifshitz 1946, Peebles 1980) solutions to the equations for the force, conservation of matter, and Poisson's equation are sought in comoving coordinates as a function of the dimensionless density contrast $\delta(\mathbf{x},t) = [\rho(\mathbf{x},t) - \overline{\rho}]/\overline{\rho}$. This variable can be expanded as a perturbation series (eg Peebles 1980), with solutions that are valid as long as $|\delta(\mathbf{x},t)| \ll 1$. When taken to the first order, i.e. linear theory, fluctuations grow at the same rate everywhere: $\delta(\mathbf{x},t) \propto D(t)$.

In the Lagrangian formalism, the variable under consideration is the displacement of a particle or fluid element from its initial position. This approach was pioneered by Zel'dovich (1970) who suggested that the relation between a particle's initial Lagrangian coordinate $\mathbf{q}$ and the final Eulerian coordinate $\mathbf{x}$ could be approximated by

$$\mathbf{x}(\mathbf{q},t) = \mathbf{q} + D(t)\boldsymbol{\Psi}(\mathbf{q}). \qquad (1)$$

Here the vector $\boldsymbol{\Psi}(\mathbf{q})$ is a time-independent function of the Lagrangian density field and D(t) is the growth factor for linear modes (see Section 2). An advantage compared to Eulerian theory is that the displacement remains finite at all times, even when $\delta(\mathbf{x},t) \to \infty$ at orbit crossing. Although displacements everywhere grow at the same rate, governed by D(t), the rate of growth of density perturbations varies as a function of position. This enables this approximation to capture some dynamical effects of the weakly non-linear regime.

In higher order Lagrangian theories (see eg Moutarde etal 1991, Bouchet et al. 1995, Catelan 1995) it is possible to treat the total displacement field as an expansion. There are however some problems with using such asymptotic series. For example Sahni and Shandarin (1995) show that the first order Lagrangian pertubation theory performs better than higher order Lagrangian theories in predicting the evolution of voids at late times. In this paper we will limit ourselves to the first order (the Zel'dovich Approximation). This has been shown to be surprisingly accurate, given its simplicity (see tests on N-body simulations by e.g. Doroshkevich et al. (1980), Efstathiou et al. (1985), Coles, Melott & Shandarin (1993)). It also has the important advantage that particles move in straight lines which enables us to use a simple computational method (see Section 2).

So far we have mentioned approximate methods which are mainly used to evolve an initial mass distribution subject to gravity forward in time. If we restrict ourselves to the growing mode of fluctuations we can in principle use them to recover the initial density. In Eulerian linear theory, the situation is trivial as the initial density is simply a scaled version of the final density. The Zel'dovich Approximation (hereafter ZA) has been recast in Eulerian coordinates and used as a tool for reconstructing the initial conditions by Nusser and Dekel (1992) and Gramman (1993a,b). The method of Nusser and Dekel is based on the requirement of momentum conservation in the ZA which is used to derive the Bernoulli equation for the evolution of the velocity potential. The approach of Gramman (1993a,b) is different, and in some respects closer to the particle trajectory based ZA as mass conservation is required. The resulting Zel'dovich-Continuity equation yields results that are more accurate than those of Nusser and Dekel, although at the expense of increased complexity.

Peebles (1989) pioneered the use of Hamilton's principle, that the action is minimised during the evolution of a collection of particles or a density field under gravity. This principle was used together with the assumption that the initial peculiar velocities of galaxies are zero in order to predict galaxy orbits in the local group (Peebles 1989) and for galaxies within 3000 kms$^{-1}$ (Shaya, Peebles and Tully 1995). Giavalisco (1993) and Susperregi and Binney (1994) also use least action in the context of an evolving mass distribution, in the latter case applied to mass smoothed onto an Eulerian grid.

Other methods of reconstructing initial conditions include that of Weinberg (1992) which involves a monotonic mapping of the final smoothed density back to the initial density on the assumption that rank order of density contrasts is preserved.

In this paper, we will use the principle of least action to describe the evolution of particle trajectories. We shall do this in the framework of the Lagrangian (particle based) ZA, which as we shall see is the least action solution for straight line particle paths. Our reasons for using this approach are as follows.

• After Eulerian linear theory, the Lagrangian ZA is probably the most used and most studied dynamical scheme, and yet it hasn't been applied to evolved density fields.
• The forward implementation of the ZA is easy to apply and computationally simple, features that we can hope to retain with an inverse ZA.
• Procedures that work in Eulerian space usually require the density field to be smoothed prior to carrying out the reconstruction. The smoothing procedure does not in general commute with the non-linear operations that are carried out on the density field. With a Lagrangian approach we can carry out the smoothing after the reconstruction.
• In principle it should be simple to make a Lagrangian procedure work in redshift space. We just need to add the predicted line-of-sight velocity for each particle to its position.

The layout of this paper is as follows. In Section 2 we examine the relationship between the principle of least action and the ZA. We describe our scheme for recovering the displacements from a final density field. In Section 3 we test the scheme on N-body simulations of CDM universes. We examine the recovery of the displacements and the prediction for the final velocity field. We then test the recovery of the initial density field. In the last two parts of this section we briefly examine how successful the scheme is when applied to a density field in redshift space and also when applied to a biased density field. In Section 4 we discuss our results further, and outline some suggestions for improvements and further work. Our conclusions are summarised in Section 5.

## 2 LEAST ACTION AND THE ZEL'DOVICH APPROXIMATION

We first consider a nonrelativistic set of particles with masses $m_i$ and comoving coordinates $\mathbf{x}_i$ in an expanding universe where non-gravitational forces can be ignored. The equations of motion can be obtained from the stationary points that result when varying the action with respect to particle trajectories. The action $\mathcal{S}$ is given by the integral of the particles' Lagrangian $\mathcal{L}$ over time, so that

$$\mathcal{S} = \int_0^t dt\, \mathcal{L} = \int_0^t dt\, [\mathcal{K} - \mathcal{W}]. \qquad (2)$$

The kinetic energy $\mathcal{K}$ is given by:

$$\mathcal{K} = \frac{1}{2} \sum m_i a^2 \dot{\mathbf{x}}_i^2 \qquad (3)$$

with $a = a(t) \equiv (1+z)^{-1}$ the universal expansion factor, which for a non-relativistic matter dominated universe obeys:

$$\begin{aligned}\frac{\dot{a}}{a} &\equiv H = H_0 \left[\Omega_0/a^3 + (1-\Omega_0-\Omega_\Lambda)/a^2 + \Omega_\Lambda\right]^{1/2} \\ \Omega_0 &= \frac{8\pi G \rho_0}{3H_0^2}\ ;\ \Omega_\Lambda = \frac{\Lambda}{3H_0^2}.\end{aligned} \qquad (4)$$

Here $H_0$ and $\rho_0$ are the present values of the Hubble constant and background density, and $\Lambda$ is the cosmological constant.

The gravitational potential energy $\mathcal{W}$ in an expanding universe can be written as an integral function of the mass autocorrelation function:

$$\mathcal{W} = -\frac{1}{2}G\ a^{-1}\ \frac{1}{V}\int dV\,\xi_2(r)/r. \tag{5}$$

The energy conservation equation for the Lagrangian can be rewritten as:

$$\frac{d}{dt}a(\mathcal{K}+\mathcal{W}) = -\mathcal{K}\dot{a} \tag{6}$$

(see e.g. Peebles 1980).

### 2.1 Straight line particle paths

We next introduce the following ansatz for the dynamics of each particle:

$$\mathbf{x}_i(t) = \mathbf{x}_i(0) + F(t)\,\mathbf{\Psi}_i \tag{7}$$

where $F(t)$ is an universal function of time and $\mathbf{\Psi}_i$ is some small initial displacement. If we assume that the initial velocities can be neglected, $\dot{\mathbf{x}}_i(0) \simeq 0$, the comoving velocity of each particle is given by:

$$\dot{\mathbf{x}}_i = \dot{F}(t)\,\mathbf{\Psi}_i = \frac{\dot{F}}{F}\Delta\mathbf{x}_i, \tag{8}$$

Particles therefore move with rectilinear trajectories (although their velocities are a function of time). In the linear regime, mass conservation requires that $\delta(\mathbf{x}_i) \propto F$, which means that $F(t) = D(t)$ is the linear growth factor:

$$D(t) = \frac{\dot{a}}{a}\int_0^a da\,\dot{a}^{-3} \tag{9}$$

(see e.g. Peebles 1980), which can be readily integrated using equation (4). In this case we recover the ZA of equation (1) in the form:

$$\dot{\mathbf{x}}_i = f\,H\,\Delta\mathbf{x}_i, \tag{10}$$

where $f \equiv (a\dot{D})/(\dot{a}D)$ is the standard linear growth factor for velocities. A reasonable approximation to $f$ is $f \simeq \Omega^{0.6}$ (Peebles 1980), although this can become slightly inaccurate if the Universe has a non zero cosmological constant (Lahav et al. 1991).

This ansatz can be used to estimate the displacement field and corresponding velocity field from a given particle distribution, as follows. The kinetic energy can be written as:

$$\mathcal{K} = \frac{1}{2}\,a^2\,\dot{F}^2\,\sum m_i\,\mathbf{\Psi}_i^2 \equiv \frac{1}{2}\,a^2\,\dot{F}^2\,\mathbf{\Psi}_0^2, \tag{11}$$

which defines the mean square particle displacement $\mathbf{\Psi}_0^2$. ¿From the energy conservation equation (6) we find:

$$\mathcal{W} = -\mathcal{K} - \frac{\mathbf{\Psi}_0^2}{2\,a}\int_0^a da\,a^2\,\dot{F}^2, \tag{12}$$

so that the action is given by:

$$S = \mathbf{\Psi}_0^2\int_0^1 \frac{da}{\dot{a}}\left[a^2\dot{F}^2 + \frac{1}{2\,a}\int_0^a da\,a^2\,\dot{F}^2\right]. \tag{13}$$

In this situation $\mathcal{K}$, $\mathcal{W}$, the Lagrangian $\mathcal{L}$ and the action $S$ are proportional to $\mathbf{\Psi}_0^2$ multiplied by a universal function of time that is fixed by the cosmology. *Thus the least action principle requires the minimum mean square particle displacement $\mathbf{\Psi}_0^2$ between the initial and final conditions.* In this paper we will use this result to estimate the set of $\mathbf{\Psi}_i$ (or $\Delta\mathbf{x}_i$) that minimize $\mathbf{\Psi}_0^2$ for a given final particle distribution $\mathbf{x}_i$ on assumption that the initial distribution is homogenous.

### 2.2 The Zel'dovich Approximation

The least action argument above holds for an arbitrary function $F(t)$ in equation (7), although the resulting solution is not necessarily a good approximation to the *exact* equations of motion. The choice $F(t) = D(t)$ produces a consistent reconstruction in the quasi-linear regime, i.e. the Zel'dovich Approximation. In general there could be further non-linear corrections, and $F(t)$ might differ from $D(t)$. Note nevertheless that one would expect corrections to be smaller for $\dot{D}/D$ than for $D$, which is why the approximation should be better than plain linear theory.

One could generalize the ZA by considering a more general expression with higher order polynomials in the linear growth function:

$$F(t) \simeq D(t) + \eta D(t)^2 + ... \tag{14}$$

which reproduces the ZA in the limit $D(t) \to 0$.[*] The first order correction to the velocities is then:

$$\dot{\mathbf{x}}_i = f\,H\,(1+\eta D)\,\Delta\mathbf{x}_i. \tag{15}$$

The value of $\eta$ can be estimated by a further minimization of the action (13) with respect to $F$, which yields:

$$\frac{d}{dt}\left[a^2\,\dot{F}\,\ddot{F} + \frac{1}{2a}\int_0^a da\,a^2\,\dot{F}\,\ddot{F}\right] = 0. \tag{16}$$

We find that for $F = D + \eta D^2$, with $D$ given by equation (9), the solution of the above equation is $\eta = 0$. Thus, the Zel'dovich Approximation $F(t) = D(t)$ provides the minimal action for straight line particle paths, which justifies, beyond the linear regime, the use of equation (10) to predict the velocities in terms of the displacements.

We have chosen to approach the problem of reconstruction of the true displacements from a purely dynamical perspective. It is also possible to appeal to more general considerations to recover the true displacements for situations in which there is no shell crossing. In Appendix A1 we show how the abscence of shell crossing in a configuration of particle trajectories determines that that configuration has the minimum mean squared displacement of all possible configurations. One consequence of this is that we could also recover the particle displacements caused by non-gravitational mechanisms for producing large-scale structure, as long as we are satisfied that shell crossing is not present. Another direct consequence is that we are able to show that the principle of least action is only strictly valid for particles moving under the ZA if there is no shell crossing.

### 2.3 Recovery of the particle displacements

To find the particle displacements we need to solve a two point boundary value problem. Our boundary conditions are the given final density distribution and the requirement that the initial density is homogenous and peculiar velocities are zero (Peebles 1989). We have already found the solution to this problem in equation (13): the mean squared displacement $\Psi_0^2$ must be minimised. As we have restricted ourselves to straight line paths, the situation is particularly simple.

---

[*] This is a particular case of the Giavalisco et al. (1993) ZA generalization, with particle trajectories restricted to be rectilinear.

**Figure 1.** The action minimisation procedure carried out on an example two dimensional particle arrangement. In each panel the initial particle positions are shown by open circles and the final positions by closed points. A solid line shows the path from each final position to its currently selected initial position. In (a) we show the initial (randomly chosen) arrangement of paths. In (b) and (c) we show an interchange of paths that is accepted by the procedure because it leads to a reduction in the action. Panel (d) shows the end state of the system once it has "cooled".

We first choose a homogenous distribution for the initial particle positions. Here we use a cartesian grid, but other types of sub-random distribution such as a glass (Baugh, Gaztañaga and Efstathiou 1995) could also be used. The situation now facing us is similar to that in the Travelling Salesperson problem, where the object is to join up a list of points with the shortest possible path. The situation here is much simpler though as each $\Psi_i$ or segment of the path is independent of all the others.

To find the displacements, we first assign a final particle position to each initial position, at random. The displacement vector $\Psi_i$ for each particle is given by the separation between the final and allocated initial position, while the action is proportional to the sum of the magnitudes of the displacements squared. The initial situation for an illustrative two dimensional density field is shown in Figure 1(a). We carry out our minimisation procedure by picking pairs of particles and interchanging their end points. If the result leads to the sum of the path lengths squared being smaller, the swap is accepted. This is shown in panels (b) and (c) of Figure 1. Particle paths are interchanged in this way until we decide to stop. We can base the decision of when to stop on the behaviour of the action. In panel (d) of Figure 1 we show the choice of particle trajectories for which the action reached a satisfactory minimum in the example case.

There are some similarities between this method and the method of simulated annealing (see e.g. Press et al. 1992) which is one way of solving the Travelling Salesperson problem. In that problem, it is necessary to accept some path swaps which increase the total length, as the path segments are not independent. If the path length is longer, whether the result is accepted depends on an artificial "temperature" for the system. This temperature is slowly reduced according to an annealing schedule so that decisions that involve large scales are made first and then, as the system "cools", finer adjustments are made to the path until convergence to a satisfactory approximation to the exact solution is reached. In our case, simulated annealing is not necessary and we can simply reject path swaps that result in an increase in the action. Our approach can therefore be described as "simulated quenching", and is free of the complications involving choice of the annealing schedule which attend simulated annealing.

## 3  TESTS ON SIMULATIONS

To check the accuracy and range of validity of our Path Interchange Zel'dovich Approximation (hereafter PIZA) method we have carried out some tests using particle spatial distributions and velocities taken from cosmological N-body simulations. In most of our tests in this paper we will deal with the idealised case of full sampling and periodic boundary conditions. Various applications of our method are checked in Sections 3.2 to 3.7.

### 3.1  Simulations used

We use simulations of two different spatially flat cold dark matter dominated Universes. One simulation is of "standard" CDM, with $\Omega_0 = 1, h = 0.5$, and the other is of low density CDM with $\Omega_0 = 0.2, h = 1$ and a cosmological constant $\Lambda = 0.8 \times 3H_0^2$. The power spectra for each model were taken from Bond & Efstathiou (1984) and Efstathiou, Bond and White (1992). Each simulation contains $10^6$ particles in a box of comoving side-length 30000 kms$^{-1}$[†] and

---

[†] Throughout the paper we use $H_0 = 100h$ km s$^{-1}$ Mpc$^{-1}$ and quote distances in units of $H_0^{-1}$, i.e. in $km/s$.

### 3.2 Minimisation of the action

We carry out the procedure outlined in Section 2.3 on each simulation, but with one additional refinement. As the size of the box is much larger than the expected maximum displacement of a particle, we speed up the algorithm, by limiting the interchange of particle paths to particles that are within a certain distance $r_{max}$. Our implementation of this idea is as follows.

(a) We first pick at random a final particle $i$ which is at position $\mathbf{x}_i$ .

(b) We then choose, also at random one of the initial grid positions, $\mathbf{x}_j(0)$, that is within $r_{max}$ of $\mathbf{x}_i$.

(c) We then assign this initial position $\mathbf{x}_j(0)$ to particle $i$ and make $\mathbf{x}_i(0)$ the initial position of particle $j$.

(d) If the sum of the path lengths squared for the new configuration is smaller than for the old, we accept the interchange and assign new labels to the initial positions, otherwise we keep the old configuration.

(e) We then return to step (a) to carry out the procedure on two more particl.es.

In order to find out when the system has reached a satisfactory state and we can stop repeating this process, we examine the behaviour of the mean squared displacement, $\Psi_0^2$. In our tests we take the value of $r_{max}$ to be 2000 kms$^{-1}$, unless stated otherwise. In panel (a) of Figure 2 we show how $\Psi_0^2$ decreases as the number of particle interchange attempts is increased. We plot results for all three of the simulations described in Section 3.1. We can see that the low density CDM model appears to be "cooling" slightly more slowly than standard CDM with the same box size and reaches a final state which has a higher value of $\Psi_0^2$. This is probably due to the fact that the low density model has more power on large scales. This means that large scale bulk flows add to the particles' average displacement. The small box standard CDM simulation reaches an even lower value of $\Psi_0^2$ due to the absence of large-scale modes. This simulation also cools more slowly, as there are 8 times as many particles to choose from within $r_{max}$ compared to the other two simulations.

In Figure 2 (b) we plot the fraction of pair interchange attempts that are successful. We can see that the slope of the curve changes slightly when $\psi_0^2$ in panel (a) starts to level off. Panel (c) shows that we can safely assume the system has reached a satisfactory solution when $\psi_0^2$ levels off. Here we plot the *rms* error on individual predicted particle velocities from our PIZA process, i.e. equation (10), against their true values in the simulations. To compare different curves, we have divided them by the dynamical factor $f = f(\Omega, \Lambda)$. We find that 500 times the number of particles is an adequate choice for the number of interchange attempts and this is the number we will use unless stated otherwise.

Our procedure is about an order of magnitude slower than competing Eulerian based systems such as the method of Nusser and Dekel (1992), although it is surprisingly fast given that it uses a technique related to simulated annealing. We have carried out our computations using an SGI Challenge computer and DEC Alpha AXP 3000 workstations, each minimisation taking 4 – 8 cpu-hours.

**Figure 2.** The evolution of various quantities during the action minimisation procedure. The horizontal axis corresponds to the number of interchange trials divided by the total number of particles. In panel (a) we show show how the mean squared displacement $\Psi_0^2$ (proportional to the action) changes. In panel (b) we plot the fraction of interchange attempts which are accepted. In panel (c) we plot the rms error on individual particle velocities divided by the dynamical factor $f$. In each case the solid and dotted lines correspond to the 30000 kms$^{-1}$ and 15000 kms$^{-1}$ box standard CDM simulations. The dashed line show the results for the 30000 kms$^{-1}$ box low density CDM model.

was run using a P$^3$M $N$-body code (Hockney and Eastwood 1981, Efstathiou et al. 1985). The mean comoving interparticle separation is therefore 300 kms$^{-1}$, of the same order as that of normal galaxies. The simulations are descibed in more detail in Croft and Efstathiou's (1994) study of galaxy cluster clustering . In order to quantify the effects of particle resolution, we have also run one additional standard CDM simulation, also with 10$^6$ particles, but in a box of side length 15000 kms$^{-1}$. This was also run with the P$^3$M code, kindly provided by G. Efstathiou. All the simulations used in Sections 3.2 to 3.6 have been evolved until the linear variance ($\sigma_8^2$) of density in spheres of radius 800 kms$^{-1}$ = 8 $h^{-1}$Mpc is equal to 1.

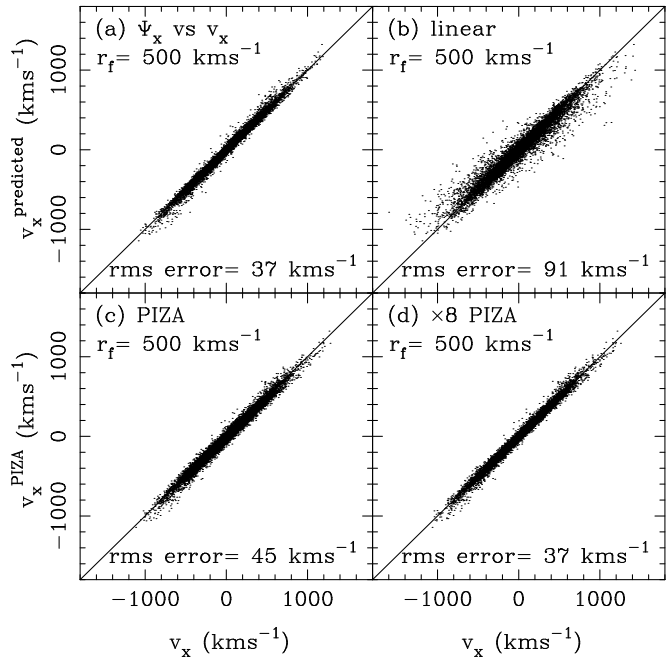

**Figure 3.** Recovery of the x component of particle displacements (panels (a) and (b)) and the smoothed displacement field (panels (d) and (d)) for the large box, Standard CDM simulation. The left hand panels show the case for one PIZA particle per simulation and in the right hand panels the predicted displacement has been obtained by an average of 8 particles (see text). In panel (a) we have added random scatter distributed like a top hat of width 20 kms$^{-1}$ to the x and y coordinates, so that the relative number of points in each band can be seen.

**Figure 4.** Recovery of the x component of the smoothed velocity field. In each case we plot $10^4$ points chosen at random from a grid of $128^3$ smoothed with a Gaussian filter with radius 500 kms$^{-1}$. In panel (a) we show the prediction of the forward ZA i.e. the velocities from a simulation plotted against the displacements taken from the simulation. In panel (b) we show the predictions of linear theory for the velocities. In panels (c) and (d) are plotted the predictions of the PIZA method for the smoothed velocity, for one PIZA particle per simulation particle and for 8 to one respectively.

### 3.3 Prediction of displacements

Once a reasonable approximation to the minimum action solution to the particle displacments has been reached, we can compare our results directly to the correct values from the simulations. In Figure 3 we plot the x components of displacements found using PIZA against the simulation values, for the case of the large box standard CDM simulation. In this section and in section 3.4, we will carry out our comparisons on quantities evaluated at the final positions of the particles.

Panel (a) shows the values for a random sample of $10^4$ individual particles. We can see that the scatter between values is large but there is also a definite correlation. More obvious is the fact that the displacements are quantised due to the initial configuration being a grid. After arriving at the individual displacements, we carry out a comparison with the smoothed displacement field. We interpolate the density onto a $128^3$ grid using a Triangle-Shaped Clouds (TSC) scheme (see Hockney and Eastwood 1981). We then also interpolate the displacement for each particle times its mass (taken to be 1) onto a grid. We smooth both fields in Fourier space with a Gaussian filter and divide the first field by the second. The values of the x-component of this displacement field at $10^4$ randomly chosen grid points are plotted in Figure 3(c). We have used a smoothing length of 500 kms$^{-1}$. The effect of smoothing is to significantly reduce the scatter and remove the quantisation. The slope of the relation remains correct.

In this trial, the initial grid used in the PIZA process is the same as the initial grid used in the "quiet start" of the simulation (see Efstathiou et al. 1985). We have also carried out the PIZA process using initial grid points displaced from the simulation values. This makes no difference to the slope or scatter about the relation except that all the predicted displacements are offset by the same factor as the initial grid offsets. This means that in the artificial situation of a simulation set up on a grid we are able to detect a bulk displacement of the mass in the box as long as it is on a scale of less than a grid cell.

Of course we are not limited to having the same number of PIZA particles as simulation particles or indeed galaxies. We have carried out the PIZA process with 8 times as many particles, assigning 8 particles to each simulation particle. In this case 200 interchange attempts per PIZA particle were necessary to reach a satisfactory minimum in the action. We then calculated the predicted displacement as the average displacement of the 8 PIZA particles associated with each simulation particle. The results are plotted in Figure 3(b). The rms error is smaller than in the one-to-one case, and there is no quantisation. The smoothed result, Figure 3(d), also has a smaller error. This means that we could use a large number of particles to carry out our smoothing in this Lagrangian way rather than on a grid. Unless we state

**Figure 5.** A slice through the density field of the 30000 kms$^{-1}$ box standard CDM simulation atfer smoothing with a 500 kms$^{-1}$ Gaussian filter. The grayscale is linear, points with an underdensity $\delta \leq -1.5$ being plotted black, and points with $\delta \geq 1.5$ being plotted white. The contours are plotted at integer values of $\delta$. (a) The initial density field. (b) The final evolved density field. (c) The initial density field reconstructed from the final particle distribution using the PIZA.

otherwise, the rest of our simulation tests will involve one PIZA particle per simulation particle.

### 3.4 Velocity Reconstruction

We now turn to the velocities predicted by the PIZA method. For comparison, we first estimate the velocity predicted by the forward ZA, equation (10), when we use the displacement actually measured between the initial position and final position of each particle in the N-body simulation (rather than the linear theory displacement).

For the Standard CDM model (which we plot in Figure 4a) we recover the correct slope for relation (10), which in this case is given by $f = 1$. For the low density model (which is not plotted), we recover a slope of f=0.41, which is the correct value for this case (see Lahav et al 1991).

For comparison, we also estimate the linear theory prediction for the velocities from the smoothed density. In linear theory the velocity $\mathbf{v}(\mathbf{r})$ is proportional to the acceleration,

$$\mathbf{v}(\mathbf{r}) = fH \, \mathbf{g}(\mathbf{r}). \tag{17}$$

In this case we have the density field on a grid with periodic boundary conditions. We solve Poisson's equation for the gravitational potential by convolving the density with the appropriate Green's function in Fourier space (see e.g. Efstathiou et al. 1985). We then find the acceleration $\mathbf{g}(\mathbf{r})$ by applying a two point finite difference operator to the potential. The predicted velocities are plotted in Figure 4(b), where we can see that the rms error is significantly worse than for the ZA.

In the lower half of Figure 4 we plot the PIZA prediction for the smoothed velocity field, obtained from the predicted displacements using equation (10). In panel (c) we have used one PIZA particle per simulation particle which gives a reasonably accurate result with about half the rms scatter of the linear theory predictions. In panel (d) we use 8 times as many PIZA particles as simulation particles. In this case we recover the full accuracy of the forward ZA. This is despite the fact that the input information is the final density rather than the initial density.

The rms error in the predicted velocities divided by $\sigma_v$, the intrinsic rms velocity dispersion forms a relative error which we can use to compare different output times. For the results shown in panels (a) to (d) of Figure 4 we find relative errors of 11%, 27%, 14% and 11% respectively. We have found in comparisons with earlier outputs of the same simulations, that as the simulation evolves, i.e. $\sigma_8$ increases, the PIZA reconstruction doesn't get any worse but linear theory does. For example, from $\sigma_8 = 0.66$ to $\sigma_8 = 1$ the PIZA reconstruction gives similar relative errors, while the linear theory prediction increases its relative error from 18% to 27%.

### 3.5 Reconstruction of the initial density field

As we have a predicted initial position for each particle, we can use this information to reconstruct the initial velocity and density fields in the simulation. We do this by assigning the PIZA predicted initial displacement for each particle to its predicted initial position, i.e. in the initial grid. The initial velocity field in the ZA, $\mathbf{v}_I = \dot{\mathbf{x}}(t_I)$ is related to the final velocity $\mathbf{v}_F = \dot{\mathbf{x}}(t_F)$ [from equation (10)] by:

$$\mathbf{v}_I = \mathbf{v}_F \frac{\dot{D}(t_I)}{\dot{D}(t_F)}. \tag{18}$$

**Figure 6.** Scatter plots of the real initial density from standard CDM simulations plotted against the reconstructed initial density using the PIZA (panels (a) to (c)) and linear theory (panels (d) to (f)). Three different smoothing Gaussian lengths have been used (the values of $r_f$ are given in each panel) and $10^4$ points taken at random from a $128^3$ grid are shown. The large box simulation was used for the left hand four plots and the $15000\,\mathrm{kms^{-1}}$ box simulation for the other two.

In this section, we are interested in testing the reconstruction of the initial density field, for which we only need to use the displacements. We assign the displacements to a $128^3$ cartesian grid as before using the TSC scheme and smooth with a Gaussian filter. The initial density field now follows from the continuity relation:

$$\delta_I(\mathbf{x}_i) = -D(t)\,\nabla\cdot\,\mathbf{\Psi}_i = -\nabla\cdot(\Delta\mathbf{x}_i) \qquad (19)$$

In Figure 5 we plot a contour plot of a slice through the smoothed density field in the large box standard CDM simulation. The Gaussian filter used in the smoothing process is of width $500\,\mathrm{kms^{-1}}$. The real initial density field used at the start of the simulation (smoothed for the plot) is shown in the left-hand panel. The amplitude of all fields has been scaled by a factor $(1+z)$ in these comparison plots. Being a Gaussian random field, this initial field has symmetry between the high and low density regions. Regions above a threshold $\delta_t$ are statistically equivalent to regions below $-\delta_t$.

The final, evolved density field (also smoothed) is shown in the centre panel. This plot also corresponds to the linear theory prediction for the initial density field (as we are using the $[1+z]$ scaling). Even though the field has been smoothed, we can see the filamentary distribution that results from gravitational instability. The symmetry between high and low density regions has now been broken. Voids have grown in size and matter has aggregated into clusters, sheets and filaments with a high relative overdensity.

In the right hand panel of Figure 5 we show the relevant slice through the PIZA reconstruction of the initial density field. The field bears a reasonable resemblance to the real initial density field of panel (a), with the agreement being best on large scales. The voids have shrunk compared to the final conditions, and the small scale features now appear closer to those in the initial density. Differences between the real and reconstructed initial density fields consist of a general lack of definition on small scales and the fact that the voids are not empty enough in the reconstruction. Both of these effects are characteristic of using the ZA, although, as with the forward ZA, many features of the density field that do result from quasi-linear processes are surprisingly well captured.

In Figure 6 we show a point by point comparison of the smoothed density fields. We use three different filters, going from large to small scales from left to right. In panel (c) we have used the higher resolution standard CDM simulation in order to make a comparison using a $250\,\mathrm{kms^{-1}}$ filter. In all the plots, the initial density is plotted on the x-axis (again scaled by $[1+z]$). The final density (corresponding to the linear theory prediction for the initial density) is plotted on the y-axis for the lower three panels and the PIZA reconstructed initial density on the upper three panels. We can again see that using the PIZA method offers a substantial improvement over linear theory. We return to a discussion of the merits and accuracy of the PIZA compared to other reconstruction methods in Section 4.

### 3.6 Reconstruction in redshift space

We can also carry out the PIZA reconstruction from the redshift space coordinates of particles. If the observer is at the origin, then for each particle, the redshift space final coordinate $\mathbf{s}_i$ is given in terms of the real space coordinate $\mathbf{x}_i$ and velocity $\dot{\mathbf{x}}_i$:

$$\mathbf{s}_i = \mathbf{x}_i + \hat{s}_i \left(\frac{\dot{\mathbf{x}}_i}{H} \cdot \hat{s}_i\right), \qquad (20)$$

where $\hat{s}_i = \mathbf{s}_i/|s_i|$ is the unit line of sight vector and $H$ the Hubble constant. Thus, in terms of the initial coordinate $\mathbf{x}_i(0)$, and displacement $\Delta \mathbf{x}_i$ we have:

$$\mathbf{s}_i = \mathbf{x}_i(0) + \Delta \mathbf{x}_i + \hat{s}_i \left(f \Delta \mathbf{x}_i \cdot \hat{s}_i\right). \qquad (21)$$

This linear set of equations can be trivially solved to obtain the three components of the displacements $\Delta \mathbf{x}_i$ in terms of the redshift and initial positions. This is all we need to carry out the PIZA reconstruction, although the results will depend on the value we assume for $f$.

To test the recovery of real space density from redshift space, we have taken the large box standard CDM simulation and added the particle's x-velocities to their x positions. In this way, as we are using the x-axis as the line of sight (the distant observer approximation).

Before we apply the PIZA process to the particle distribution, we must first deal with the "fingers of God" caused by virialised galaxy clusters. These objects are the result of strongly non-linear gravitational collapse on small scales, which we cannot hope to follow with the ZA. Here we use the approach of Gramman, Cen & Gott (1994) in choosing to collapse the fingers along the line of sight. We employ a friends-of-friends groupfinder written and first used by Davis et al. (1985), which joins together into groups particles separated by less than a certain linking length. We have modified the groupfinder so that the linking length is different for separations parallel to and perpendicular to the line of sight. To collapse the fingers, we place all particles in a group at the average redshift of the group particles. The linking lengths are chosen by estimating the correlation function of the particles as a function of separation across and along the line line of sight. For the correct linking lengths, the correlation function should be approximately symmetric on small scales ($r < 5\ h^{-1}\mathrm{Mpc}$) (see Gramman et al 1994). Here we use linking lengths of 0.15 and 1.5 times the mean interparticle separation for distances across and along the line of sight respectively.

We then carry out the reconstruction process on the redshift space particle distribution. The line-of-sight distance between the initial and final positions in redshift space is larger than in real space (twice as large if $\Omega = 1$). We therefore use a larger search distance $r_{max} = 4000$ kms$^{-1}$ in the direction parallel to the line of sight. If we assume a value for $f$ and hence $\Omega$, the PIZA process yields predicted displacements and velocities for the particles. We then use the predicted velocities to reconstruct the final real space density distribution. In principle, we could choose the correct value of $f$ by picking the value which reproduces the isotropy of the real-space correlation function, this time on intermediate to large scales ($r > 5\ h^{-1}\mathrm{Mpc}$). This would be one way of using the ZA to recover $\Omega$ from redshift space distortions of clustering (for another way see e.g. Fisher and Nusser 1995). We will discuss this further in Section 4. In the meantime, as we are using the standard CDM simulation, we take $f = 1$.

We again assign the particles to a $128^3$ grid using the TSC scheme and then smooth the density with a Gaussian filter. When this smoothing process is carried out, we make sure that the fingers of God are kept compressed and that their predicted velocity is given by the average of the PIZA velocities of the particles they contain. We have already seen (in Figure 3a) that because of relatively large separation of the particles, the scatter in velocities of individual particles is rather large. This scatter in velocities will translate into a blurring of the redshift space density reconstruction on small scales. To improve the results, we could reduce the scatter by using more particles. In order to keep matters simple in this preliminary test, we instead use a filter size large enough to smooth over the small scale scatter.

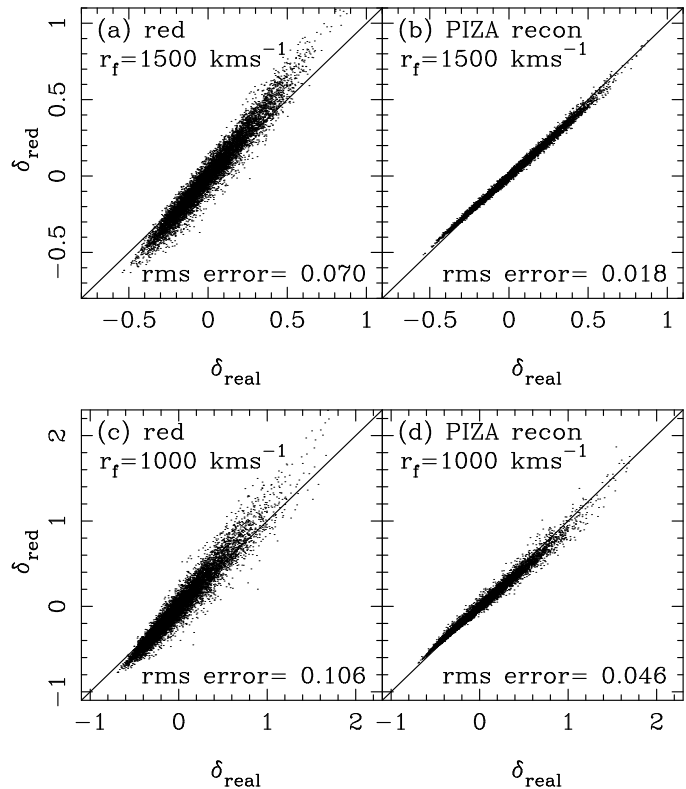

**Figure 7.** Recovery of the real space final density from redshift space final density in the 30000 kms$^{-1}$ box standard CDM simulation. (a) and (c) show the redshift space density (y-axis) plotted against real space density (x-axis). (b) and (d) show the recovered real space density using the PIZA predicted velocities (y-axis) against the true real space density (x-axis). Two different Gaussian smoothing filters were used, 1500 kms$^{-1}$ (top) and 1000 kms$^{-1}$ (bottom), and a random sample of $10^4$ grid points are plotted.

In Figure 7 we show the results for two different filter sizes, $r_f = 1500$ kms$^{-1}$ (top panels) and $r_f = 1000$ kms$^{-1}$ (bottom panels). In all the panels, the real space density is plotted on the x-axis. In (a) and (c), we have plotted the redshift space density from the simulation on the y-axis. In (b) and (d) we have plotted the PIZA reconstructed final real space density on the y-axis. We can see from panels (a) and (c) that transformation into redshift space increases the relative density contrast as well as introducing some scatter. The PIZA reconstruction restores the correct slope, even in regions with a density contrast $\delta > 1.5$, as well as reducing the scatter. As expected, the reconstruction works best after filtering on large scales. If we just compress the fingers of God without carrying the reconstruction, we see only minimal improvement in the scatter, in line with the results of Gramman, Cen & Gott (1994).

**Figure 8.** Reconstuction of the initial density field in the 30000 kms$^{-1}$ box standard CDM simulation from redshift space. On the x-axis we plot the smoothed initial density field in the simulation and on the y-axis we plot the smoothed PIZA reconstructed initial density field. The amplitude of both fields has been scaled by (1+z) and a random sample of 10$^4$ grid points is plotted in each panel.

We have also carried out the recovery of real space density in the low density CDM simulation. Random virial velocities are not as important in this model, and we are able to recover the real space density with similar accuracy to the standard CDM trial even without compressing the fingers of God.

Since the final real space density can be recovered fairly well, at least when smoothed on large scales, we now test the reconstruction of the initial density field from the redshift space final density. After applying the PIZA process, we have to correct the particle final positions into real space using their predicted velocities. After this, the process is identical to that detailed in Section 3.3. The initial density and the recovered initial density for the standard CDM model (after using filters of sizes 1500 kms$^{-1}$ and 1000 kms$^{-1}$) are shown in Figure 8. Comparing Figure 8b and Figure 6a we can see that the scatter in the reconstruction has increased by a factor of $\sim$ 2.5 with respect to the result obtained in real space. The reconstruction is still reasonable, though, and there may exist ways to reduce this error, which we will discuss in Section 4.

### 3.7 Application to biased simulations

When we apply a dynamical scheme to galaxy redshift data rather than to the mass taken directly from a simulation, we must take into account that the large scale galaxy distribution might not be good tracer of the mass. In dynamical schemes based on linear theory, it is common to invoke the linear biasing model for the relationship between galaxies and mass. In this model, it is assumed that if the galaxy density constrast $\delta_g(\mathbf{x})$ is smoothed on sufficiently large scales then the mass density contrast is given by $\delta_m(\mathbf{x}) = \delta_g(\mathbf{x})/b$. This model is probably reasonable on large scales (see e.g. Bardeen et al. 1986, Kaufmann et al. 1995), and has been extensively used (see e.g. Dekel & Rees 1987, Strauss et al. 1992, Fisher et al. 1994). In these studies, there is a degeneracy between the bias parameter $b$ and the dynamical factor $f$ in equation (10). If we wish to break this degeneracy, we need to examine the situation on smaller scales, using dynamical schemes which are valid in the quasi-linear regime. Unfortunately, on scales small enough that we expect non-linearity in the gravitational evolution of mass, we should also expect to find non-linearity in the relationship between galaxies and mass (see e.g. Fry & Gaztañaga 1993) and maybe even scale dependent biasing (see Gaztañaga & Frieman 1994, Weinberg 1995) . If the biasing process can be approximated by a local transformation, $\delta_g = F[\delta_m]$, then in Eulerian schemes it is possible to use a model to map the galaxy density to mass density before carrying out a reconstruction.

With a particle-based scheme such as that presented here, it is not obvious what we should do. For example, if galaxies are more clustered than mass then we should somehow add particles to the underdense regions. However even in the case of linear biasing, there does not appear to be any simple prescription for "de-biasing" the galaxy distribution using particles.

To test the possible effects of biasing, we apply a prescription to select a population of galaxy particles that have different clustering properties from the underlying mass. For this test we use the standard CDM 30000 kms$^{-1}$ box simulation and choose galaxy particles as follows. For each particle we find the distance to the 20th nearest particle and then calculate the overdensity inside a sphere of this radius. If the resulting fluctuation is above a certain critical value (in this case 0.5) we add the particle to the list of galaxies. These galaxy particles are preferentially selected in high density regions and are more clustered than the underlying mass. We also add a fraction (30%) of the non-peak particles to the list of galaxies to make up for the fact that the finite resolution of the simulations means that power is missing on small scales which might have conceivably caused a few galaxies to be formed in less overdense regions. We use a simulation output which has $\sigma_8 = 0.66$ for the mass. When we apply the biasing procedure we find 89$^3$ galaxy particles and the resulting distribution has a rms fluctuation is tial of being more accurate than linear theory, even when used on a biased density distribution. If we apply the PIZA method to the unbiased distribution, i.e. with $\sigma_8 = 0.66$, we predict the velocity field with even more accuracy (the rms percentage error is 14%). This is shown in Figure 9b. If there were some way of "de-biasing" the galaxy distribution, this could done in conjunction with a least squares fit to the velocity field. Unfortunately there does not seem to be a simple way of doing this, at least with the PIZA.

## 4 DISCUSSION

As far as comparison with the Eulerian ZA reconstruction methods is concerned, panel (a) and (b) of Figure 6 should be compared with the bottom panels of Figures 2(a) and 2(b) in Nusser and Dekel (1992). Our formulation of the ZA gives a substantial improvement over the Zel'dovich Bernoulli reconstruction. The PIZA has similar accuracy to the Zel'dovich continuity approach of Gramman (1993a,b); this is reasonable as both approaches are based on the requirement of mass conservation.

Other methods for initial density reconstruction (e.g. those of Nusser and Dekel 1992 and Gramman 1993a) tend to require that redshift density be converted into real space density prior to the reconstruction, a process which usually requires an iterative technique (see e.g. Gramman et al. 1994). For the Gaussian-mapping process of Weinberg (1992), the fact that the slope in Figure 7a is not 1 does not matter, as it is the rank ordering of densities which is important. The scatter seen about the relationship will mean increased errors in the reconstruction though. As we have shown in Section 3.6 the PIZA method can handle redshift space density fields directly. The method will need some modification to deal with observed samples of galaxies with a selection function as the rocket effect (Kaiser 1987) will affect the reconstructed fields. For example, linear theory could be used initially to recover the real space selectio function.

In principle, we could also use the method to measure $\Omega$, as the correlation function of the reconstructed real space density should be isotropic. Some tests of this idea are needed, and it is uncertain how the potentially complex relationship between galaxies and mass will affect the results. One advantage of a dynamical reconstruction from redshift space to find $\Omega$ is that we would not need to use the small angle approximation. This approximation (that the observer is sufficiently distant that pair separations long the line of sight are parallel) is usually used and may lead to systematic errors for redshift surveys with large sky coverage (see e.g. Hamilton and Culhane 1995).

An additional advantage of our particle-based scheme is the fact that it is numerically stable even though no smoothing is used before it is applied. Eulerian schemes usually require the application of differential operators to the density field, so that smoothing is required before starting the reconstruction. This is potentially problematic as smoothing must (first at least) be carried out in redshift space, which will mean an additional loss of information on small scales, particularly if particle velocities are large. Also, the smoothing and non-linear dynamical operators do not commute. This means that a self-consistent treatment of smoothing

**Figure 9.** Recovery of the x component of the smoothed velocity field from biased and unbiased simulations with $\sigma_8 = 0.66$ for the underlying mass and $\sigma_8 = 1.0$ for the biased particles. In each case we plot $10^4$ points chosen at random from a grid of $128^3$ smoothed with a Gaussian filter with radius 500 kms$^{-1}$. The lefthand panels show the prediction for the velocity field from PIZA (a) and linear theory (c) when applied to the biased simulations. The prediction has been scaled by one over the linear bias factor (b=1.5) in both these panels. Panel (b) shows the PIZA prediction for the velocity of the unbiased simulation.

spheres of $8h^{-1}$Mpc of $\sigma_8^b = 1.0$. We define a linear bias factor, $b$ as:

$$b \equiv \frac{\sigma_8^b(biased)}{\sigma_8(mass)}, \qquad (22)$$

so that in this case we have $b = 1.5$.

We apply the PIZA method to this simulated galaxy distribution in the same way that we have previously done with the mass. We take the predicted velocities assigned to a grid and smoothed with a Gaussian filter as before and compare them with the real smoothed velocities of the galaxy particles. The results are shown in panel (a) of Figure 9 , where we have scaled the prediction with the inverse of the linear bias factor $b$.

Unlike in the case of linear theory, it is not obvious that the linear bias factor should apply when relating the predicted and observed velocities. We do expect particles to have to travel farther in the biased simulation, but it is interesting that this results in recovery of the same slope as linear theory.

We have also used linear theory to predict the velocities from the accelerations of the biased particles. The results, also scaled with $b$, are being shown in panel (c) of Figure 9. We can see that the scatter from in the linear theory plot is again worse than for the PIZA prediction (the rms percentage error is 26% as opposed to 18% for PIZA). This result shows that our reconstruction method still has the poten-is not possible in general in Eulerian schemes, which therefore sometimes require the application of empirically derived smoothing corrections to the final results.

## 5 SUMMARY AND CONCLUSIONS

If the trajectories of particles or fluid elements are restricted to be straight lines (as in the Zel'dovich Approximation) then the principle of least action leads to the simple constraint that the sum of the path lengths squared is minimised. This fact can be used in conjunction with a numerical scheme to recover the displacement field from a final particle distribution. This scheme is both intuitively simple and easy to implement. We have carried out tests of the scheme on N-body simulation outputs of CDM universes and find that when reconstructing the velocity field the results have the same relatively small errors as an application of the forward ZA. It is therefore natural to consider the scheme to be a numerical inversion of the Lagrangian ZA. Although the scheme is particle based, it is possible to minimise the action of several million particles. This means that it should be possible to assign several particles to each galaxy when using observational data in the form of galaxy redshift surveys. We should therefore be able to treat galaxy merging in a limited way, and in any case we have shown here that the use of more than one particle per galaxy increases the accuracy of the reconstruction. Our scheme can work in redshift space, and can also be used to recover the real space density from redshift space density.

As with other reconstruction methods our uncertainty as to the relationship between galaxies and mass is one of the fundamental limits to the accuracy of results. A study of dynamics and galaxy biasing on scales small enough to have some non-linearity needs a tool that can follow the gravitational evolution of the mass with reasonable accuracy on these scales. This is true of the PIZA method, which we have shown to be substantially more accurate than linear theory.

### Acknowledgements

We would like to thank George Efstathiou for supplying us with a copy of the P$^3$M code and groupfinder. We also thank the anonymous referee for useful comments and suggestions. RACC is supported by NASA Astrophysical Theory Grants NAG5-2864 and NAG5-3111 and would like to thank David Weinberg for useful discussions and comments on the manuscript. EG thanks the Astrophysics group in Oxford for their hospitality, and acknowledges support from CSIC, DGICYT (Spain), project PB93-0035 and CIRIT (Generalitat de Catalunya), grant GR94-8001.

# APPENDIX A1: SHELL CROSSING AND MINIMUM MEAN SQUARED DISPLACEMENT

It is possible that the configuration of displacements $d_i$ which connects pairwise the initial $r_i(0)$ and final positions $r_i$ could have a minimum mean squared value with respect to any other possible pair configuration, regardless of the least action principle or the Zel'dovich approximation.

This is trivially the case when the displacements are very small compared to the interparticle separation. To see this we consider the set of all possible pair configurations, $\{r_i(0), s_i\}$ where $s_i = r_j$ is given by a set of $j = 1, .., n$ indices that are permutations of the $i = 1, .., n$ indices. Let us call $\Psi_T^2 = \sum d_i^2$ the true mean square displacement. For any other configuration we have $\Psi^2 \equiv \sum_i (r_i(0) - s_i)^2$ which gives:

$$\Psi^2 = \Psi_T^2 + 2\sum_i d_i(r_i - s_i) + \sum_i (r_i - s_i)^2. \quad (A1)$$

When the displacements are small, $d_i \to 0$, we have:

$$\Psi^2 = \Psi_T^2 + \sum (r_i - s_i)^2. \quad (A2)$$

so that for any other configuration $\Psi^2 > \Psi_T^2$, and therefore $\Psi_T^2$ is the minimum of all possible configurations.

The above argument cannot be used in general when the displacements $d_i$ are large. The term $2\sum d_i(r_i - s_i)$ does not vanish by isotropy because in general the configuration that minimizes $\Psi^2$ does not have to be isotropic with respect to the true configuration.

We will now show that when there is no shell crossing ($nsc$) in the trajectories of a configuration, its mean squared displacement is the minimum of all possible pair configurations.

Let us first define *shell crossing (sc)* and *mean squared displacement*.

DEFINITION 1. Given two trajectories connecting the position $r_i(0)$ with $r_i$ and $r_j(0)$ with $r_j$, we will say that there is **shell crossing** between these trajectories whenever:

$$[r_i - r_i(0)]^2 + [r_j - r_j(0)]^2 > [r_i - r_j(0)]^2 + [r_j - r_i(0)]^2. \quad (A3)$$

That is, the total path of the actual trajectories is larger than the path of the crossed trajectories, where we swap the final (or initial) positions.

DEFINITION 2. Given two sets of positions $r_i(0)$ with $i = 1,...n$ and $r_j$ with $j = 1,...n$, consider the group of all pair configurations given by the set of all permutations of the index $j$, which we label $j_i = \{j_1,..,j_n\}$. For each configuration $\{j_i\}$ connecting positions $r_i(0)$ with $r_{j_i}$, we define the **mean square displacement** $\Psi^2$ as:

$$\Psi^2 \equiv \frac{1}{n}\sum_i [r_{j_i} - r_i(0)]^2. \quad (A4)$$

We can now prove the following theorem.

THEOREM. A pair configuration has a minimum **mean square displacement** if and only if there is **no shell crossing** ($nsc$) in this configuration.

Let us label the one particular configuration we are interested in by $T_i$, with mean squared displacement: $\Psi_T^2$. We can formally express this THEOREM in the following notation:

$$nsc\ \{T_i\} \iff \Psi_T^2 \le \Psi^2 \quad \forall \{j_i\} \quad (A5)$$

This is equivalent to saying that there is shell crossing ($sc$) in $\{T_i\}$ if and only if there is some configuration $\{j_i\}$ with $\Psi^2$ for which $\Psi_T^2 > \Psi^2$:

$$sc\ \{T_i\} \iff \exists \{j_i\} \quad \Psi^2 < \Psi_T^2. \quad (A6)$$

The proof of the THEOREM has therefore two steps. We first prove that:

$$sc\ \{T_i\} \implies \exists \{j_i\} \quad \Psi^2 < \Psi_T^2. \quad (A7)$$

This can be shown as follows. If we just swap (permute) the pairs in the sum of $\Psi_T^2$ of $\{T_i\}$ where there is shell crossing, it follows from the definition of $sc$ that the permuted configuration will have a smaller value of $\Psi^2$, and therefore (A7) is true. We now show that the other direction holds also:

$$\exists \{j_i\} \quad \Psi_T^2 > \Psi^2 \implies sc\ \{T_i\}. \quad (A8)$$

This can be shown as follows. Consider the configuration $\{j_i\}$ for which the above is true. Group arbitrarily the indices $i = 1, ..., n$ in pairs and call the members of each pair $\{k1, k2\}$. We will have:

$$\Psi^2 = \sum_{k1}[r_{j_{k1}} - r_{k1}(0)]^2 + [r_{j_{k2}} - r_{k2}(0)]^2 \quad (A9)$$

where $j_{k1}$ and $j_{k2}$ are the indices of the final positions associated with $k1$ and $k2$ in the configuration $\{j_i\}$, for which $\Psi^2 < \Psi_T^2$. We do the same grouping for the $\{T_i\}$ configuration. As all terms above are positive and $\Psi^2 < \Psi_T^2$, it follows that there is at least one term in the sum above which is smaller than the corresponding term in the sum for the $\{T_i\}$ configuration. That is, there is shell crossing for at least the pair corresponding to this term and therefore (A8) is true.

An immediate consequence of this theorem is that there are many situations for which one can recover the displacements (by looking for the minimum in $\Psi^2$) and which might have nothing to do with gravity or the Zel'dovich Approximation (ZA). This is similar to the situation which occurs with the relation $\delta \sim \nabla v$ which also applies to non-gravitational systems (Babul et al. 1995). In our case, however, it does not apply to the velocities, which in general do not have to be proportional to the displacements. If we decide arbitrarily that they do, we would still have to appeal to gravitational dynamics and the Zel'dovich Approximation for the proportionality factor.

Another direct consequence is the following corollary

COROLLARY The least action principle for particles moving under gravity in an expanding universe requires that there is no shell crossing as long as particles move in straight lines (that is under the ZA).

We have shown in section 2 that the configuration which results from application of the least action principle has the minimum mean square displacement and therefore, under the above THEOREM this means that there is $nsc$. Another way of saying this is that the ZA can only be valid as far as there is $nsc$. The appeal of the approach we take in Section 2 is that for straight line paths we have both a dynamical approach (ZA) and a reconstruction method (minimum $\Psi^2$) which are consistent with each other and which yield accurate results with large enough smoothing.